\documentclass[preprint,times]{elsarticle}

\usepackage{float}
\usepackage{hyperref}
\usepackage{setspace}
\usepackage{xcolor}\usepackage{caption}
\usepackage{amssymb}

\usepackage[figuresright]{rotating}

\begin{document}

\doublespacing

\begin{frontmatter}

\title{Tidal Rock Grinding as a Source of H$_2$ on Enceladus}

\author[harvard]{Karin I. \"Oberg}
\author[ethz]{Cara Magnabosco}
\author[cambridge]{Nicholas J. Tosca}

\address[harvard]{Center for Astrophysics | Harvard \& Smithsonian, 60 Garden St., Cambridge, MA 02138, USA}
\address[ethz]{Department of Earth and Planetary Sciences, ETH Zurich, Sonneggstrasse 5, Zurich, CH-8092, Switzerland}
\address[cambridge]{Department of Earth Sciences, University of Cambridge, Cambridge, CB2 3EQ, United Kingdom}

\begin{abstract}

The Solar System hosts multiple potentially habitable environments, including the subsurface ocean beneath the icy crust of Saturn’s moon Enceladus. This ocean’s composition is unusually well constrained thanks to Cassini’s observations of Enceladus’s south polar plume during multiple flybys. Among the plume’s more surprising components is molecular hydrogen (H$_2$), detected in trace amounts. The presence of observable amounts of H$_2$ is intriguing not only because it is difficult to explain through conventional geochemical processes, but also because it could serve as both an energy source for life and a driver of prebiotic organic chemistry.
In this study, we explore whether tidally induced rock grinding within Enceladus’s core could account for the observed H$_2$. Laboratory experiments show that H$_2$ can be efficiently produced when freshly fractured rock reacts with water. Using these experimentally determined production efficiencies, we estimate H$_2$ generation rates as a function of the fraction of tidal energy dissipated through rock grinding.
Our results suggest that tidal grinding could plausibly produce the observed levels of H$_2$, with instantaneous production rates potentially exceeding those from radiolysis or serpentinization. However, sustaining such production over geological timescales would require efficient healing of silicate surfaces in the core to allow repeated grinding. Without such healing, tidally induced rock grinding may instead lead to episodic bursts of chemical activity lasting up to millions of years—potentially sufficient to initiate new prebiotic pathways. This transient mechanism would complement the longer-term, lower-energy contributions from serpentinization (over hundreds of millions of years) and radiolysis (over billions of years).
\end{abstract}


\end{frontmatter}

\section{Introduction} \label{sec:intro}

Planetary habitability is commonly evaluated in terms of access to liquid water, organic and inorganic reactants, and sources of energy that can power and potentially direct the chemistry towards life, as well as sustain ecosystems once life exists. Within this context, the icy moons in the outer solar system are of particular interest \cite{Lammer2009,McKay08}. Several of them show clear evidence of subterranean water oceans \cite{Carr1998,Hansen06}. These oceans are likely rich in organics based on the organic content of their presumptive building blocks--carbonaceous chondrites and comet-like bodies--and the similar sublimation properties of water and common small organic cometary molecules like methanol, ethanol, dimethyl ether, and formic acid  \cite{Oberg09d}. Furthermore, experiments on the sublimation properties of water and organic ice mixtures show that even the more volatile organic content is retained, with the partial exception of the hypervolatile CH$_4$, during ice heating as long as the water ice is preserved \cite{Collings04}. Recent D/H measurements in the icy crusts of Saturnian moons appear to show that relatively pristine water ice was indeed directly incorporated into these solar system bodies \cite{Brown25}, and we should hence expect subterranean oceans in the Saturnian statelites to contain large abundances of organic material. In the case of Saturn's tiny moon Enceladus, these expectations have been spectacularly confirmed by mass spectrometric observations of relatively large abundances of organic molecules in its plume as well as in icy particles in the Saturn E ring, which is sustained by Enceledus \cite{Waite2006,Postberg2018,Peter2024}.

One of the most exciting discoveries related to the Enceladus plume is the presence of H$_2$ \cite{Waite17}; Based on Cassini-INMS measurements, \cite{Waite17} derived a H$_2$/H$_2$O ratio corresponding to a H$_2$ release rate of $1-5\times10^9$  mol H$_2$ year$^{-1}$. By comparison, the Earth's H$_2$ release estimates are about two orders of magnitude higher \cite[e.g.][]{bach2003iron,cannat2010serpentinization,lollar2014contribution}, and when taking into account the different sizes of Enceladus and Earth (Enceladus is 60,000 times smaller), the Enceladus flux is quite remarkable. On Earth, the use of H$_2$ as an electron donor constitutes one of the phylogenetically ancient forms of metabolism \cite{Raymann2015}, and perhaps equally important, H$_2$ production in aqueous environments could enable  prebiotic chemistry to develop the organic complexity on which origins of life depend \cite{Preiner2020,Kaur2024}. The presence of H$_2$ in the Enceladus plume and, by inference, its ocean, may thus hint at a rich prebiotic chemistry that could under the right conditions proceed to produce life. 

However, explaining the observed H$_2$ in the Enceladus plume has proven challenging. It is highly unlikely that it is due to the outgassing of primordial H$_2$, as H$_2$ capture was not expected during Enceladus's formation, either by gas accretion or clathrate formation, and neither the ocean nor the ice shell would be able to store sufficient H$_2$ over geological time \cite{Waite17}. Multiple {\it in situ} H$_2$ formation processes have instead been considered, out of which radiolytic and hydrothermal production in Enceladus core have received most attention \cite{Vance2016,Bouquet17}. \cite{Bouquet17} used a model of radiolysis from unstable isotopes in ordinary chondrites to calculate the production rate of H$_2$. The resulting estimates are only consistent with the Enceladus plume measurements in the distant past, while the more recent production is at most $\sim$10$^8$ mol H$_2$ year$^{-1}$, an order of magnitude lower than required by observations. Instead serpentinization is the currently favored explanation; \cite{Bouquet17} find it to be 10x more efficient than radiolysis. The presence of substantial serpentinization is also supported by the observation of nano-silicate grains in Enceladus plume -- a telltale sign of rock-water interactions \cite{Hsu2015}. Models suggest, however, that the current H$_2$ production through this channel may barely exceed the observed H$_2$ escape rate \cite{Waite17}. Furthermore, serpentinization could at most maintain the observed H$_2$ escape rate over 500 Myrs, and a recent study suggests that ``either the hydrothermal activity has developed recently on Enceladus, or alternative processes (pyrolysis of insoluble organic matter, microbial
activity) must be tested to explain the observed H$_2$ flux in Enceladus' plume." \cite{Daval22}. Indeed, a recent Bayesian analysis found that methanogenesis may be the most likely explanation when also taking into account the methane measurements in the Enceladus plume \cite{Affholder2021}.

We propose that a possible alternative process to explain the observed H$_2$ is a group of reactions that could occur between ground silicate and water in the Enceladus core. On Earth, radical reactions on fault surfaces may generate substantial amounts of H$_2$ during earthquake faulting~\cite{Hirose2012}, glacial comminution~\cite{telling2015rock}, and tidal erosion~\cite{he2021abiotic}. Here, silicon-oxygen bonds of silicate mineral surfaces are broken and form surface radicals that split water and produce reactive oxygen species (ROS) and H radicals, which recombine to form H$_2$, O$_2$ and H$_2$O$_2$~\cite{He23}. 
\cite{parkes2019rock} has shown that such mechanoradically-derived H$_2$  can sustain methanogenic microbial communities, supporting its possible importance for subterranean life. Furthermore, the co-produced oxidizing products, H$_2$O$_2$ and O$_2$, have been proposed to serve as electron acceptors for prebiotic chemistry as well as microorganisms \cite{Stone2022,He23}. 
Given that tidal energy dissipation in the rocky core of Enceladus may similarly result in faulting and grinding of silicate minerals in contact with water, surface oxygen radical reactions with water may have a substantial impact on the chemical environment in the Enceladus ocean, and contribute to the observed H$_2$ in the Enceladus plume. 

In this paper we use recent experimental data on H$_2$ production from aqueous interactions with ground up rocks together with estimates of tidal dissipation in the Enceladus core to calculate a range of plausible H$_2$ production rates in the Enceladus core, the transport of this H$_2$ to the plume, and the possible cumulative yield over Enceladus's life time (\S2). We compare these instantaneous production rate estimates and cumulative yields to those of other proposed H$_2$ formation mechanisms on Enceladus, and discuss how the different mechanisms may impact both H$_2$ production and the overall Enceladus prebiotic chemistry over different time scales (\S3). After discussing limitations with the current modeling approach, we offer some concluding remarks and possible future directions in the final section (\S4).

\begin{figure}[ht]
    \centering
    \includegraphics[width= 0.5\textwidth]{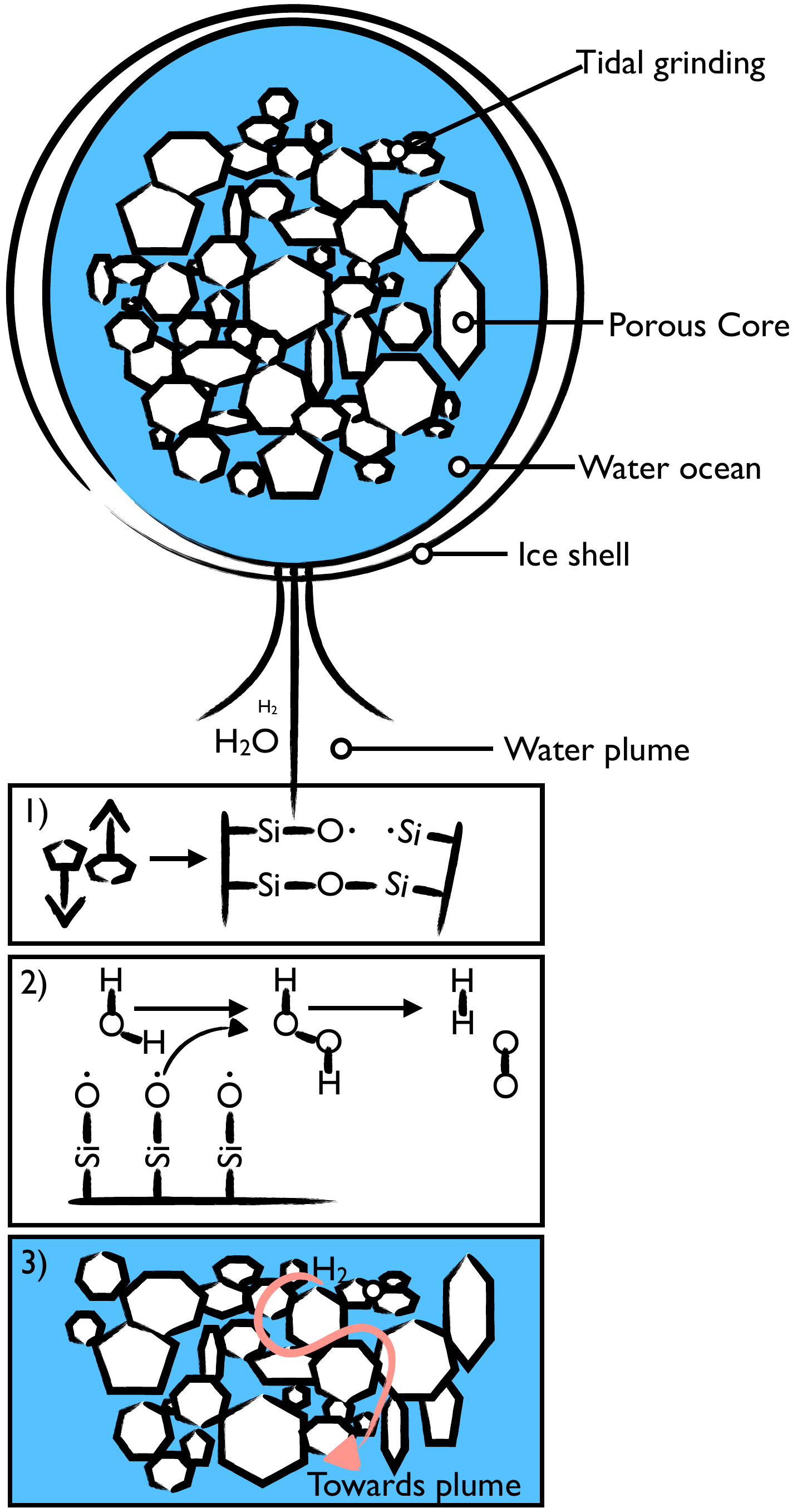}
    \caption{A cartoon representation of the structure and relevant processes of Enceladus, showcasing its icy surface, internal ocean and rocky core as well as the three steps required for tidally induced rock grinding in the core to explain H$_2$ observations in the plume.}
    \label{fig:enceladus}
\end{figure}

\section{Tidally Driven \texorpdfstring{H$_2$}{H2} Production and Transport in the Enceladus Core}\label{sec:model}

The question of whether reactions between silicate-surface oxygen radicals and water is a plausible explanation for the observed H$_2$ in Enceladus's plume can be split up into five sub-questions: 1) At what rate are oxygen radicals produced in Enceladus's core? 2) What fraction of the surface-oxygen radicals react with water to produce H$_2$ molecules? 3) How efficiently are these H$_2$ molecules transported into Enceladus's ocean and further to the south pole plume reservoir? 4) What chemical sinks may limit H$_2$ (and O$_2$) transport to the plume? 5) Over what time periods can the overall process be sustained? These questions are illustrated in Fig. \ref{fig:enceladus} and treated sequentially in the following sub-sections.

\subsection{Production Rate of Silicate Surface Radicals \label{ros-production}}

To calculate the production rate of reactive oxygen on silicate surfaces requires an estimate of the amount of tidal energy that is dissipated in Enceladus's core. This is not a well-established number due to uncertainties both about the orbital dynamics of the Saturnian moons over geological time, and regarding where and how tidal energy is dissipated in the interior of Enceladus. Early estimates suggested that the total amount of tidal energy dissipation in Enceladus should be quite low, on the order of 1 GW \cite{Meyer07}, with negligible dissipation occurring in the core. Such a low value seems at odds with the presence of a liquid subterranean ocean on Enceladus as well as the measured heat loss, both of which require more than 20 GW of tidal energy dissipation to be sustained over geological time \cite{Choblet17}. This discrepancy appears largely resolved, however, in the more recent literature. New measurements of the orbital dynamics of the Saturnian moons, and new models of Enceladus's orbital evolution and structure have resulted in considerably higher estimates of the total tidal dissipation loss. The new estimates of 10-55 GW are more consistent with estimates of the present-day heat loss  \cite{Choblet17,Lainey20, rovira2022tides, Nimmo23}. We adopt 20 GW as our fiducial value, but acknowledge that this number may need future revision.

A second unknown is the fraction of the total tidal energy that is dissipated in Enceladus's rocky core. Early models, which assumed a rigid, compact silicate core, predicted negligible tidal dissipation in the core \cite{Nimmo07}, but this has since been revised due to new gravity measurements that revealed a low density core, compared to what would be expected if the core was solid rock \cite{Iess14}. Other evidence for a porous core comes from Enceladus's librational modes, and the realization that Enceladus's core likely never experienced sufficient heat and/or pressure to compact it \cite[e.g.][]{Thomas16}. In a porous core tidal energy dissipation could be efficient \cite{Roberts15,Choblet17}, and for the purpose of this paper we assume, consistent with recent models, that at least 50\% of the tidal energy, or 10 GW, is deposited within the Enceladus porous core. 

We next consider how tidal energy dissipation may result in the production of fresh silicate rock; on Earth a variety of geologic processes have been reported to generate silicate surface radicals, and many of these may also occur within the core of Enceladus. Generally speaking, tidal dissipation occurs through material deformation, which is instantiated through a number of mechanisms, including slips, faults, and dislocation creep. These could all result in the production of fresh rock surfaces. In addition, one could imagine more indirect tidal energy dissipation impacts on the rocky core composition through rock cracking from cooling-heating cycles due to expansion during ice formation in the core, followed by tidally induced melting. Which mechanism dominates, especially in the kind of porous core that is proposed for Enceladus, is largely unknown, complicating efforts to estimate the production rate of Si-O$\cdot$ surface radicals. However, considering that the rheological properties of the Enceladus core are typically modelled as a viscoelastic~\cite[e.g.][]{ross1989viscoelastic,Roberts15} or poroviscoelastic~\cite[e.g.][]{liao2020heat,rovira2022tides} water filled silicate matrix, we expect that creep, abrasion and shearing are the dominant processes responsible for generating Si-O$\cdot$ surface radicals during long-term tidal deformation. Importantly, these deformation processes are largely aseismic in nature and do not require high magnitude seismic events to generate Si-O$\cdot$ surface radicals and are likely to be more similar to what is observed during glacial comminution~\cite{telling2015rock} and tidal erosion~\cite{he2021abiotic}.  

In environments where the rock composition, morphology and geological processes are well understood, the production of fresh silicate surfaces could in theory be calculated directly taking into account slip velocities and other relevant factors. This is not the case for the Enceladus core, however, where there are large uncertainties involved in translating between existing models of Enceladus's core processes and fresh rock production through any of the proposed deformation processes. 
Therefore, for the purpose of this study, we simply assume that 10\% of the tidal dissipation energy in the core results in the grinding of rocks and breaking silicate Si-O bonds to form Si-O$\cdot$ surface radicals, but also consider H$_2$ production rates for higher and lower values below. This value will certainly require revision, once more detailed models of tidal dissipation in the Enceladus core exist. Assuming a total tidal dissipation energy of 10~GW, a 10\% bond-breaking efficiency implies that 1~GW is available for creating fresh silicate surfaces in the Enceladus core.

Given this assumption, the number of exposed O radicals in the core can be estimated by comparing the fraction of tidal energy dissipation rate that goes in to breaking Si-O bonds (1 GW) using an average Si-O bond strength. This value depends on the precise mineral, and may vary between 470 and 720 kJ/mol \cite{Huhn2022}. Here we adopt an approximate value of $\sim$600 kJ/mole. This implies the production of oxygen radicals on silicate surfaces of ${\rm 10^9 \:J\: s^{-1} / 6\times10^5\: J\: mole^{-1}\sim 2\times10^3 \: moles \: s^{-1}}$ or $6\times10^{10}$ moles year$^{-1}$, or a factor of two higher if the cleavage is instead between two oxygen atoms resulting in two oxygen radicals per broken bond.

\subsection{Production Rate of \texorpdfstring{H$_2$}{H2} \label{h2-production}}

In this section we consider what fraction of the silicate oxygen radical sites  may result in H$_2$ production. There are several experiments that find measurable production of H$_2$ and other relevant species (H$_2$O$_2$ and O$_2$) following exposure of freshly ground rock to water \cite{Hurowitz2007,Hirose2012,Stone2022,He23}, demonstrating that qualitatively the process is robust to different experimental conditions. For the purpose of this study we rely mainly on \cite{Stone2022}, whose recent experiments provide quantitative yields across a range of environments.

\cite{Stone2022} measured both the production of Si-O radical sites and the production rate of H$_2$ at different temperatures. They estimate that at the initiation of their experiments, they had 13.0 --14.4 $\mu$mol g$^{-1}$ of silicon radical sites (and by extension O radical sites), and observed a maximum H$_2$ production of 3 $\mu$mol g$^{-1}$, i.e. the maximum proportion between silicate reactive oxygen sites and H$_2$ production is $\sim$20\%. Most of their experiments were carried out at $60^\circ$C and above, with activity peaking between 80 and $104^\circ$C for all considered crushed rock types (granite, basalt, and  peridotite). Crushed granite appears to produce the most H$_2$, while the mineral with the lowest H$_2$ production, basalt, appears to have an activity of about 30\% compared to granite.

As indicated above, the H$_2$ formation rate increases with temperature up to $104^\circ$C. While the temperature in Enceladus's core is  relatively unconstrained by either observations or models, some limits arise from the boiling and freezing points at the ocean floor. The former is $\sim290^\circ$C \cite{Glein2018}, and reactions at higher temperatures  than considered in experiments are hence plausible. On the other hand, there may not be hydrothermal activity on Enceladus, and it is possible that the ocean may instead be close to freezing. In addition to the warmer experiments listed above, 
  \cite{Stone2022} carried out some preliminary experiments at lower temperatures, 0 and 30$^\circ$C. In these experiments, very little H$_2$ was initially produced, but after one week of monitoring, the cumulative H$_2$ production rate was comparable to the experiment run of $60^\circ$C. This suggests that while the H$_2$ production rate is highly sensitive to the water temperature, the total conversion rate of silicate oxygen radicals into H$_2$ molecules is not. We therefore use a yield of 20\% of H$_2$ production per O-site as a our fiducial value, but also consider lower and higher values of 2\% and 50\% respectively.

\begin{figure}[ht]
    \centering
    \includegraphics[width= 0.8\textwidth]{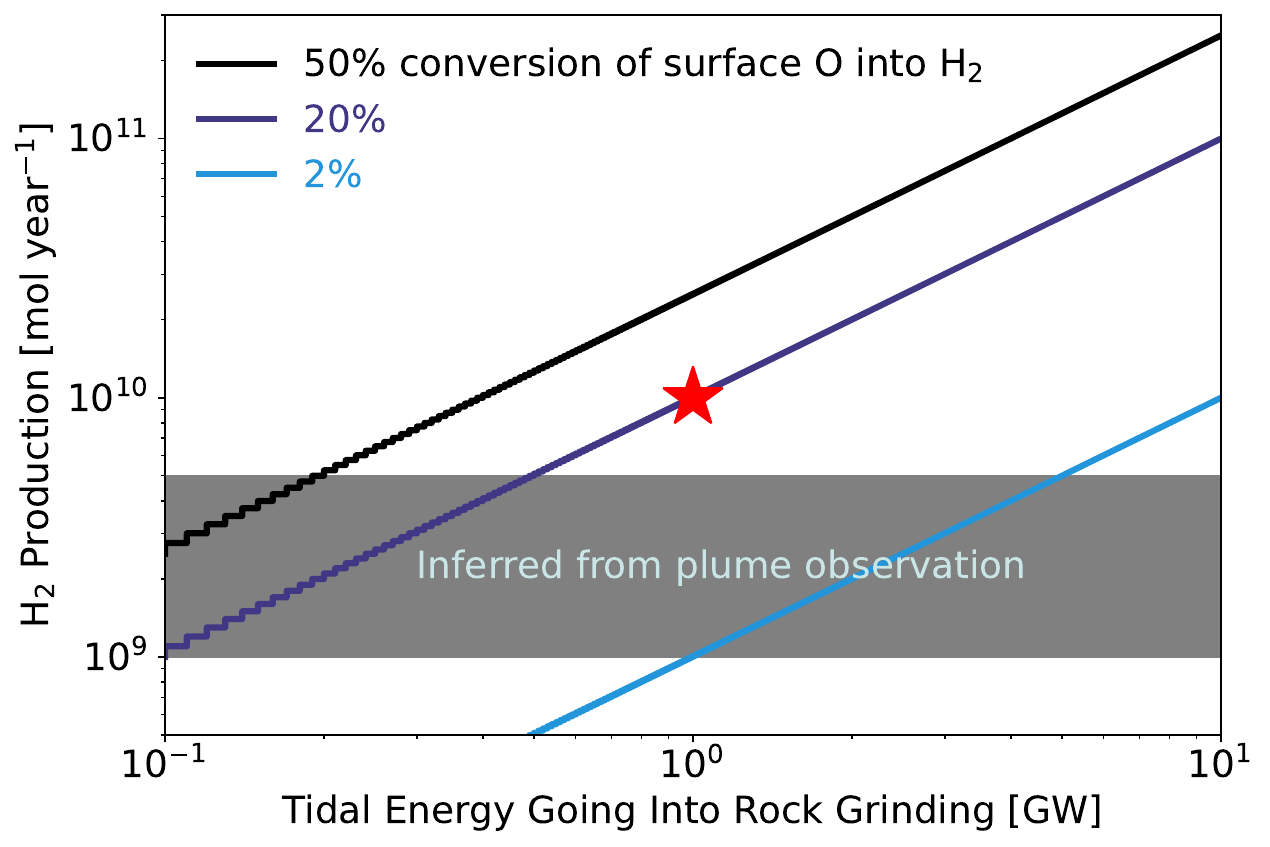}
    \caption{H$_2$ production rate in Enceladus core for different assumptions on tidal rock grinding efficiencies and conversion rates of exposed oxygen radicals on fresh silicate surface into H$_2$. The grey region marks the lower limit of H$_2$ production from H$_2$ plume measurements. The red star marks the fiducial estimate.}
    \label{fig:H2}
\end{figure}

Figure \ref{fig:H2} summarizes the expected H$_2$ production in the Enceladus core for a range of rock grinding and H$_2$ production efficiencies assuming steady state. If the conversion of oxygen radical sites on fresh rock faces is 20\% (our fiducial value), sufficient H$_2$ is produced to explain plume measurements of $\sim$10$^9$ mol per year, even if only 0.5\% (0.1 GW) of the tidal energy goes into breaking rock silicate bonds. By contrast a low H$_2$ production efficiency, which seems more consistent with some of the older experiments \cite{Hirose2012}, would require a much larger proportion of tidal energy (5--50\%) being dissipated through silicate bond breaking to still be consistent with observed plume H$_2$ fluxes.

\subsection{\texorpdfstring{H$_2$}{H2} Transport \label{transport}}

We next consider how diffusion may reduce the flux of H$_2$ molecules out of the Enceladus core compared to the calculated production flux above. Using a simple set of fluid transport considerations, we re-cast the bulk H$_2$ production rate in terms of a surface area-normalised flux of H$_2$ to solution ($J$; mol m$^{-2}$ sec$^{-1}$); this is required to evaluate diffusive transport in porous media such as the Enceladus core. Expressing the bulk H$_2$ production rate in terms of a surface area-normalised flux requires constraints on the concentration of Si-O radicals per unit surface area, which are available from experimental measurements. This also requires internally-consistent assumptions related to the total surface area and thus particle size in the core. \color{black} We estimate the combined flux resulting from production at silicate mineral surfaces and from diffusion, as a function of the Darcy velocity of the fluid ($q$; m sec$^{-1}$). At high rates of fluid flow across silicate mineral surfaces, the combined flux will be limited by the production rate of Si radicals prescribed above, while lower fluid flow rates will result in reduced, diffusion-limited fluxes.

Similar to above, we use the experimental results from \cite{Stone2022} together with the density of basalt to estimate the concentration of Si radicals per unit area of silicate material and find $1.3 \times 10^{-4}$ mol Si-radicals m$^{-2}$, which we assume is independent of particle size. Adopting a production rate of $6 \times 10^{10}$ moles of Si radicals per year, and a 20\% efficiency of H$_{2}$ production, yields an integrated H$_2$ production across the entire core of 380 mol H$_{2}$ sec$^{-1}$, corresponding to the fiducial value in Fig \ref{fig:H2}. Assuming constant concentration of Si radicals per unit area then yields the rate by which ``reactive surface area'' is generated if the particle size distribution in the core is known, which produces Si radicals and ultimately H$_{2}$ at $1.4 \times 10^{7}$ m$^{2}$ sec$^{-1}$. To express these values as a surface-area normalized total reaction flux, and to facilitate comparison with diffusive fluxes, requires assuming a total surface area present in the core.  We estimate total surface area from the total core volume at an average density equal to basalt at three different average particle sizes representative of particle sizes in protoplanetary disks and chondritic material: 1~$\mu$m--1~mm. This range also incorporates the typical particle size from \cite{Stone2022} following rock grinding. This calculation results in the maximum H$_2$ production flux (that is, a flux that is limited by the availability of Si radical surface sites), which is highest when reactive surface area is largest relative to total surface area. The results below scale linearly with this number, and can hence be readily adjusted as the surface radical production estimate becomes solidified. The diffusive flux, $J_{D}$ (mol m$^{-2}$ sec$^{-1}$), through a static boundary layer positioned at the silicate mineral surface is:

\begin{equation}
    J_{D} = k_{D}(c_{s} - c)
\end{equation}

where $k_{D}$ is a mass transfer coefficient (m sec$^{-1}$), and $(c_{s} - c)$ is the difference in concentration between the silicate surface and bulk solution (mol m$^{-3}$). The flux of H$_2$ from the reaction itself is:

\begin{equation}
    J_{R} = k_{R}(c_{eq} - c_{s})
\end{equation}

where $k_{R}$ is a mass transfer coefficient (m sec$^{-1}$) corresponding to the reaction producing H$_2$ at silicate mineral surfaces and $(c_{eq} - c)$ is the difference in concentration between the equilibrium H$_2$ concentration, and the concentration in the bulk solution (mol m$^{-3}$). Because the equilibrium concentration is unknown, and because our production estimates (described above) constrain the maximum reaction fluxes ($J_{R}$) subject to assumptions related to total surface area availability, we set $(c_{eq} - c)$ equal to 1 mol m$^{-3}$ which ensures that $J_{R}$ simply equates to the total reaction fluxes estimated above. 

At steady state, $J_{R} = J_{D}$ which provides an expression for the surface concentration $c_{s}$, which can be substituted into either of the above equations, resulting in a combined reaction and diffusion flux:

\begin{equation}
    J = \frac{k_{R}k_{D}}{k_{R} + k_{D}}(c_{eq} - c)
\end{equation}

The mass transfer coefficient associated with diffusion across a static boundary layer surrounding a packed bed of spherical particles ($k_{D}$) can be estimated with the following empirical expression \cite{Cussler2009}:

\begin{equation}
    k_{D} = 1.17q\left(\frac{Lq}{v}\right)^{-0.42}\left( \frac{v}{D} \right)^{2/3}
\end{equation}

where $D$ is the diffusion coefficient of H$_{2}$ in H$_{2}$O at 298~K (5.11 $\times10^{-9}$ m$^2$ sec$^{-1}$), $L$ is the particle diameter (m), $q$ is the Darcy velocity of the fluid (m sec$^{-1}$) and $v$ is the kinematic viscosity of H$_2$O (8.917 $\times10^{-7}$ m$^2$ sec$^{-1}$). We note that the temperature dependencies of $D$ and $v$ across 0 to 200$^{o}$C introduce negligible variation from the 298~K case \cite{Kerkache.2025}.

\begin{figure}[ht]
    \centering    \includegraphics[width=0.8\textwidth]{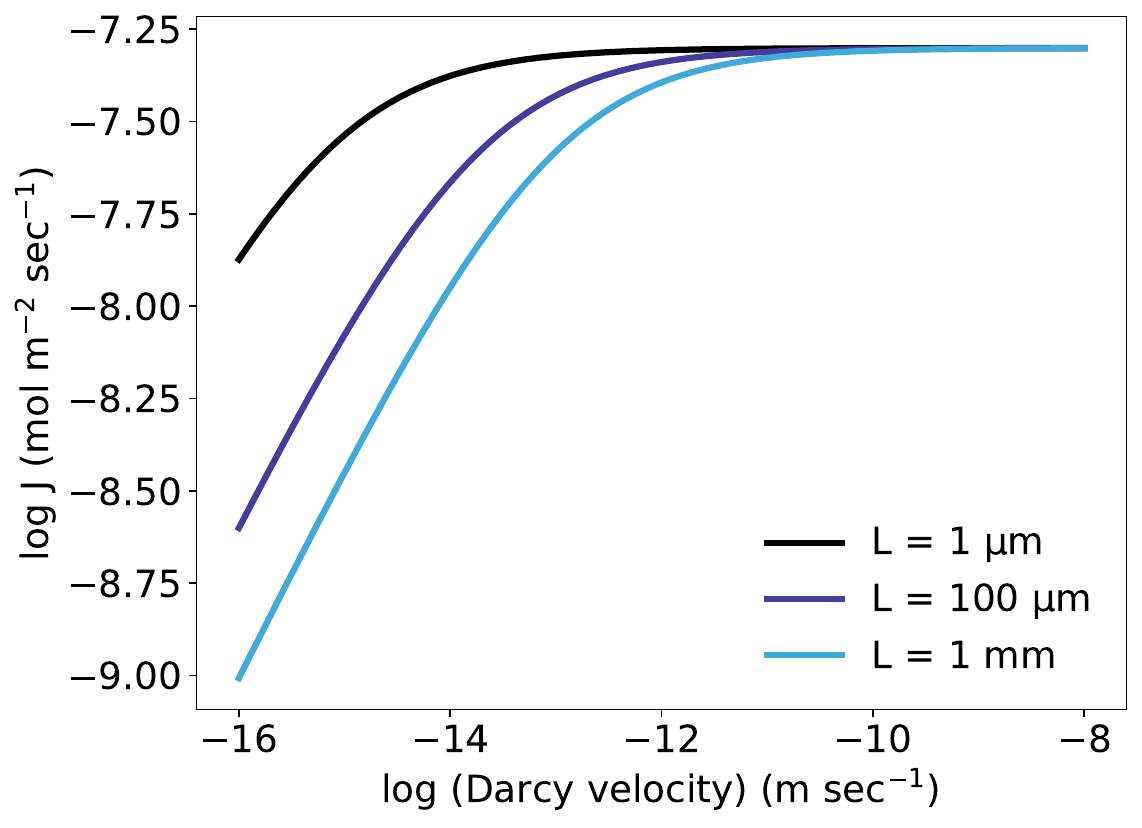}
    \caption{The estimated H$_2$ flux per unit area of exposed silicate in the core as a function of average core grain size, assuming an inherently small grain size population, and the associated diffusion rate, where the latter is governed by the Darcy velocity.}
    \label{fig:transport}
\end{figure}

Figure \ref{fig:transport} shows that over a large range of fluid velocities, the combined H$_2$ flux is limited by production of H$_2$ at silicate mineral surfaces. Diffusion of H$_2$ through H$_2$O begins to limit the combined H$_2$ flux at relatively low fluid velocities (below 10$^{-10}$ m sec$^{-1}$). For comparison, \cite{rovira2022tides} model calculations of tidal dissipation in the Enceladus core support fluid velocities of up to $2\times10^{-5}$ m sec$^{-1}$. 

This analysis shows that when transport is diffusion-limited, the H$_2$ production rate largely sets the expected flux in the plume. However, convection or other types of fluid flow may be more important in H$_{2}$ transport than diffusion if the time (or length) scale is sufficiently long \cite{Cussler2009}. We can examine the lengths at which transport of a dissolved constituent via diffusion and via advection are equal given a transport time, $t$. The advective transport time over distance $L$ is $t_{adv} = L/q$ where q is the Darcy velocity. The diffusive transport time over the same distance is $t_{diff} = L^2/2D$. Setting these two times equal gives $L = 2D/q$, where diffusion and advection contribute equally to transport. At 298~K and $q = 2\times10^{-5}$ m sec$^{-1}$, $L$ is approximately 500 $\mu$m. At distances shorter than this, diffusion should dominate transport, and at distances larger than this, advection should dominate. The adopted core grain sizes above are either in the diffusive regime, or at the boundary, validating the above calculation. If the characteristic grain size is larger, the transport rates will need to be re-calculated considering advection. In the meantime, it  seems reasonable to adopt the H$_2$ production flux at the silicate surface as the H$_2$ flux into the Enceladus ocean.

\subsection{H$_2$ and O$_2$ Lifetimes}

While the previous calculation shows that H$_2$ can be efficiently transported to the ocean, it does not take into account H$_2$ chemical reactivity, which may be a limiting factor. Furthermore, O$_2$ is co-produced with H$_2$, but not observed in the Enceladus plume, and we must hence consider the chemical fates of both O$_2$(aq) and H$_2$(aq) to assess the plausbility of the proposed H$_2$ production scenario on Enceladus. 

\begin{figure}[ht]
    \centering    \includegraphics[width=0.8\textwidth]{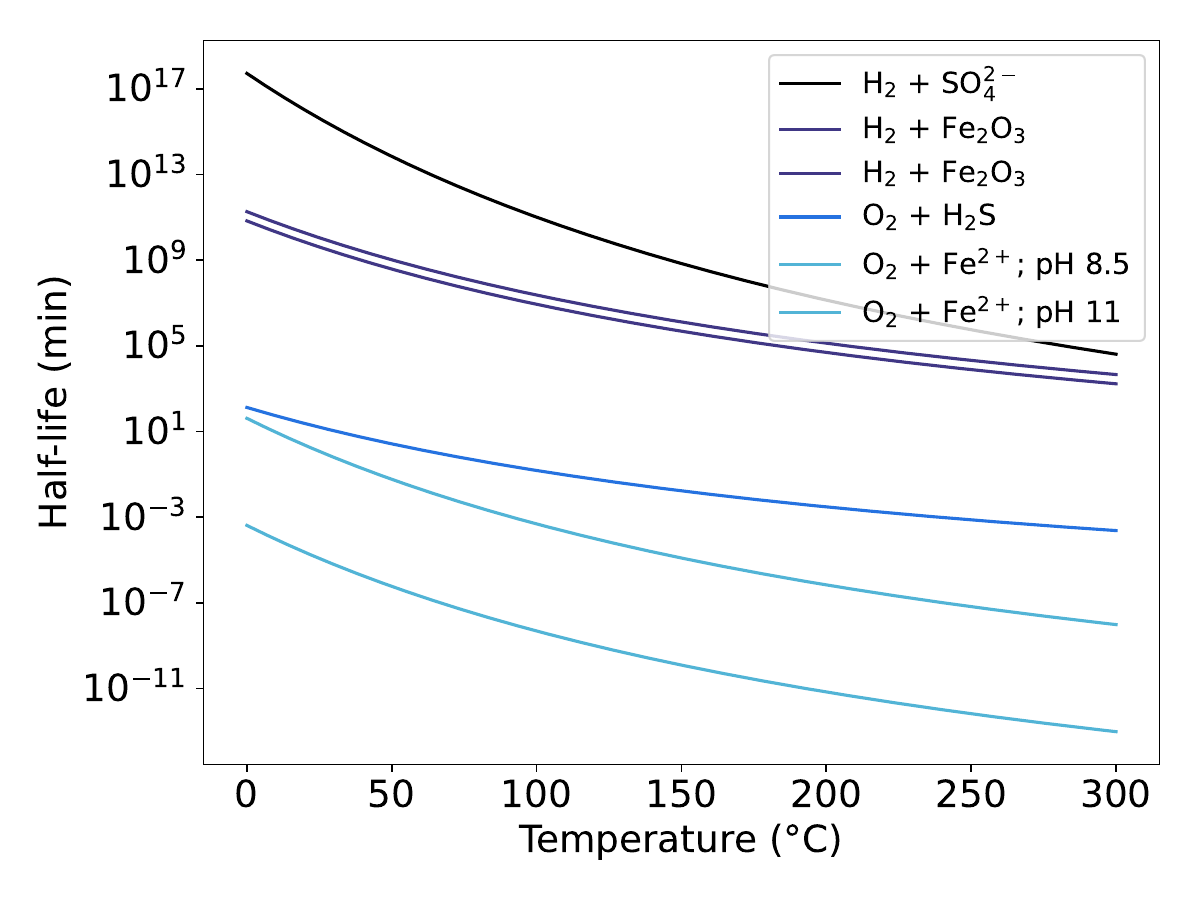}
    \caption{The estimated half-lives of O$_{2}(aq)$ and H$_{2}(aq)$, in the presence of geochemically-relevant reductants and oxidants, as a function of temperature. Reactions consuming H$_{2}(aq)$ include: $\rm3Fe_{2}O_{3} + H_{2}(aq) \rightarrow 2Fe_{3}O_{4} + H_{2}O$ (with rate constants estimated using data reported in \cite{Seyfried.1987} and activation energy constrained from \cite{Pineau.2006}), and $\rm 4H_{2}(aq) + SO_{4}^{2-} + 2H^{+} \rightarrow H_{2}S(aq) + 4H_{2}O$ \cite{Truche.2009}. Reactions consuming O$_{2}(aq)$ include: $\rm 4Fe^{2+} + O_{2}(aq) + 4H^{+} \rightarrow 4Fe^{3+} + 2H_{2}O$ ([Fe$^{2+}$] = 1e$^{-6}$ mol kg$^{-1}$, pH 8.5-11; \cite{Stumm.1996}), and $\rm H_{2}S(aq) + 2O_{2}(aq) \rightarrow SO_{4}^{2-} + 2H^{+}$, which under the alkaline conditions at Enceladus becomes $\rm HS^-(aq) + 2O_{2}(aq) \rightarrow SO_{4}^{2-} + H^{+}$} ([$H_{2}S(aq)$] = 1e$^{-5}$ mol kg$^{-1}$; \cite{Stumm.1996,Zhang.1993}). Values for pH, [Fe$^{2+}$], and [$\rm H_{2}S(aq)$] are chosen to be consistent with geochemical constraints imposed by Cassini data \cite{Xu.20252fh}.
    \label{fig:chemistry}
\end{figure}

Generally, whether O$_2$(aq) and/or H$_2$(aq) are likely to reach measurable concentrations in water depends on the their rates of reaction with available reactants. We therefore compiled the reaction rates of O$_2$(aq) and H$_2$(aq) with geochemically-relevant reductants and oxidants in aqueous media at a wide range of temperature. These data are most clearly expressed as the half-life of O$_2$(aq) or H$_2$(aq) (the time required for the reaction to deplete their concentrations to half of their initial values). Figure \ref{fig:chemistry} shows a significant difference between the aqueous reactivity of O$_2$(aq) and H$_2$(aq); H$_2$(aq) is comparatively unreactive relative to O$_2$(aq). The latter is rapidly consumed (on timescales of seconds to hours), while the rates of H$_2$(aq) consumption by oxidized minerals (e.g., hematite) or oxidized S compounds (e.g., SO$_4^{2-}$ and/or other compounds of intermediate oxidation state) are low (half lives of years to millions of years) across a wide range in temperature.

 Notably, the most rapid reaction we identified that could consume H$_2$(aq) is a reaction with O$_2$(aq) in solution to form H$_2$O (H$_{2}$(aq) + 1/2O$_{2}$(aq) = H$_{2}$O \cite{Foustoukos.2011}), which could consume H$_2$ faster than any of the reactions considered in Fig. \ref{fig:chemistry} if there is significant build-up of O$_2$(aq). However, because, the rates of reaction between O$_2$(aq) and Fe$^{2+}$(aq) or H$_2$S(aq) (in the form HS$^-$ under Enceladus's alkaline conditions) are considerably faster than the H$_{2}$(aq) + 1/2O$_{2}$(aq) reaction across the estimated pH range of the Enceladus ocean, such a build-up is unlikely.  Together, these results indicate that O$_2$(aq) is likely to be efficiently consumed in the Enceladus ocean, whereas H$_2$(aq) is likely to reach higher concentrations owing to rates of consumption comparable to geological timescales.

These results generally make intuitive sense. The sluggish reactivity of H$_2$(aq) is well known to chemists and geochemists who choose to use H$_2$ gas to maintain anoxic atmospheres in the laboratory because its reaction rates with oxidizing compounds are so slow across wide ranges in temperature; it is also well known to those who study the dissolved gas concentrations of aqueous fluids circulating through Earth’s crust (such as modern subseafloor hydrothermal fluids), which are invariably poor in O$_2$(aq) relative to H$_2$(aq) \cite{VonDamm.1995}. Taking into account the differing chemistry of H$_2$(aq) and O$_2$(aq) it is hence reasonable to expect that the H$_2$ flux in the Enceladus plume is largely set by its production flux, while the O$_2$ flux would be diminished by many orders of magnitude.

One possible issue with the proposed O$_2$ consumption reactions is that in an unbuffered system and over geologial timescales they would lead to substantial SO$_4^{2-}$ accumulation and ocean acidification, at odds with the alkaline ocean environment inferred from Cassini measurements (e.g. \cite{Xu.20252fh}); SO$_4^{2-}$ is produced when O$_2$ is consumed by H$_2$S, and following the initial reaction between O$_2$ and Fe$^{2+}$, Fe$^{3+}$(aq) will react with water following $\rm 4Fe^{3+} + 8H_2O \rightarrow 4FeOOH + 12H^+$. However, the Enceladus rock-ocean system is likely characterized by an enormous buffering capacity, which would limit the impact of the cumulative O$_2$ consumption on ocean composition and pH. We discuss this in more detail below after considering what other constraints exist on the possible cumulative H$_2$ production from rock-grinding.

\subsection{Cumulative \texorpdfstring{H$_2$}{H2} Production and Sustainability Over Geological Time \label{sec:model-cum}}

Given that the H$_2$ flux is mainly set by H$_2$ production, we finally turn to assessing the possible H$_2$ production over the lifetime of Enceladus. To estimate the maximum cumulative H$_2$ production from rock grinding we assume the extreme where the total silicate mass of Enceladus participates in H$_2$ formation following the empirical results of \cite{Stone2022}. In this study, granite, peridotite, and basalt were all crushed to a mean grain size of $\sim20$ $\mu$m, resulting in the production of $\sim$3~$\mu$mol H$_2$ per gram of rock. Assuming $30\%$ porosity of Enceladus's $2.1\times10^7$~km$^3$ core, a core composition similar to ordinary chondrites (65 wt \% SiO$_2$, density of 3 g cm$^{-3}$), and H$_2$ production rate of 3~$\mu$mol H$_2$ per gram SiO$_2$, grinding down the entirety of the Enceladus core would produce $\sim9\times10^{16}$ moles of H$_2$. This implies that this production channel could sustain the minimum H$_2$ generation  rate of $1\times10^9$ mol per year implied by plume measurements for $\sim10^8$ years, i.e. much shorter time than the age of the solar system.  This result indicates that in the absence of rock healing processes that restore Si-O-Si linkages, tidally induced rock grinding could not sustain substantial H$_2$ production over geological time.   

On Earth cracks and fissures in rocks can self-repair or heal through mineral precipitation and chemical reactions which occur at silicate mineral surfaces (i.e., the very rapid condensation of two terminal Si-OH surface groups to generate a Si-O-Si linkage and H$_{2}$O; \cite{Pelmenschikov.2001,Casey.2003}). The efficiency of such processes depends on the mineral, the size of the cracks, the composition of the aqueous solution, as well as the temperature and pressure environment \cite[e.g][]{Brantley1990,Brantut2015}. Without experiments designed to mimic the conditions in the Enceladus core it is difficult to estimate the importance of such processes. We note, however, that  on Earth, water is a key healing agent because of its ability to transport dissolved mineral constituents, and then to precipitate them in cracks and pores \cite[e.g.][]{Rutter1976,Richard2015}. It is hence plausible that water plays a similar role in Enceladus, and perhaps quite effectively so if water permeates the whole core. Any hydrothermal activity at the core-ocean interface may also enable higher-temperature healing processes. 

If efficient rock healing processes are present in the Enceladus core, the maximum cumulative yield of H$_2$ from rock grinding can be estimated based on the total chemical potential of the core, i.e. the number of O-O and Si-O bonds in the silicate matrix. We estimate the number of SiO$_2$ units in the core to be $\sim5\times10^{20}$ moles, using the core volume, porosity, and compositional assumptions listed above. If 10\% of SiO$_2$ units produce an O radical during the lifetime of Enceladus and 20\% of these radicals yield a H$_2$ molecule (our fiducial yield assumption), the cumulative H$_2$ production would be $\sim1\times10^{19}$ moles, i.e. sufficient to sustain a production rate of  $\sim1\times10^{9}$ moles per year over geological time. 

We now return to whether we can use the present-day Enceladus ocean composition and alkalinity to put additional constraints on the possible cumulative H$_2$ production. A cumulative H$_2$ production of $\sim1\times10^{19}$ moles would imply a similar production of O$_2$, and therefore of protons following the O$_2$ consumption chemistry introduced above. In addition, if oxidation of reduced sulfur is a major removal path of O$_2$, the total production of SO$_4^{2-}$ could be inconsistent with the lack of reported sulfate ions in plume particles, whose analysis has for example yielded detections of phosphate \cite{Postberg2023}, if the sulfate concentration is proportional to the cumulative O$_2$ production. We first note that SO$_4^{2-}$ production is only a concern if H$_2$-based consumption of O$_2$ dominates, while our calculations show that reactions with Fe$^{2+}$ are faster. Second, what sulfate is produced should effectively be removed by reactions to form more reduced forms of sulfur, whose abundances are then limited by Fe-sulfide precipitation \cite{Xu.20252fh}; the estimated high equilibrium partial pressure of H$_2$(g), imposes an extreme thermodynamic drive for the chemical reduction of sulfate to reduced species, especially in the presence of several possible reductants such as Fe$^{2+}$, CH$_4$, other simple organic compounds, and reduced sulfur itself. Instead we consider the ocean pH to be a more strict constraint on cumulative H$_2$ production since all proposed O$_2$ consumption reactions produce at least one proton per O$_2$ consumed. The impact of these protons on ocean pH and composition depends ultimately on the ocean buffering capacity, and related, which buffering system dominates. 

To estimate a baseline buffering capacity, we first note that a simple Henderson-Hasselbalch treatment, in which added protons are assumed to titrate only the dissolved bicarbonate pool at fixed total inorganic carbon and alkalinity, would substantially underestimate the buffering capacity of the Enceladus ocean, since the ocean is in active contact with a hydrated silicate core. In such a rock–water system, pH is instead controlled by total alkalinity coupled to carbonate–silicate mineral equilibria; added protons drive carbonate mineral precipitation rather than being neutralized by dissolved HCO$_3^-$ alone, and the ocean can therefore draw on a mineral reservoir that is orders of magnitude larger than the aqueous carbonate inventory. We accordingly estimate the buffering capacity in two steps: first from the dissolved carbonate–water system alone, as a conservative lower bound, and then including the additional contribution from silicate-mineral buffering at the ocean–core interface.
In the first approach, we use the nominal Enceladus ocean composition suggested by \cite{Glein2025}, and take pH = 10.6, and an ocean volume of $\approx3 \times 10^{19}$ L, which results in a dissolved CO$_{2}$ activity of $\approx10^{-6.2}$ \cite{Glein2025} and a resulting buffering capacity of $2 \times 10^{19}$ total equivalents of alkalinity\footnote{We note that the individual carbonate species concentrations tabulated in \cite{Glein2025} (Table A1) do not close the alkalinity budget when inserted directly into the standard definition, yielding only $\sim 0.1$~eq~kg$^{-1}$, because those values were constructed assuming a total dissolved inorganic carbon rather than being derived self-consistently from pH and CO$_2$ activity. Since pH and CO$_2$ activity uniquely determine carbonate speciation and alkalinity at a given temperature, the self-consistent alkalinity implied by the inferred pH ($\sim 10.6$) and $a_{\mathrm{CO_2}} \sim 10^{-6.2}$ is substantially larger, $A_T \sim 0.6$--$1$~eq~kg$^{-1}$, and it is this alkalinity that controls the ocean's long-term buffering capacity.}. Thus, assuming a yearly proton production rate of $2\times10^{10}$ moles, pH buffering due only to the composition and size of Enceladus's ocean leads to the following rate of pH change:

\begin{equation}
    \rm \Delta pH\;\textrm{/ year} \approx \frac{2\times 10^{10}\;\textrm{mol/year}}{2\times 10^{19}\;\textrm{mol/pH unit}} \approx 1 \times 10^{-9},
\end{equation}

which equates to $\approx$1 pH unit per billion years. However, the presence of silicate minerals at the ocean-core interface provide buffering capacity that is far greater still. Assuming that hydrous silicates such as serpentine and/or talc represent the bulk of the core, H$^{+}$ would also be consumed in the following reaction:

\begin{equation}
    \rm \textrm{Serpentine} + 1.5H^{+} + 1.5HCO_{3}^{-} \rightarrow 3H_{2}O + 1.5 MgCO_{3} + 0.5 \textrm{Talc}
\end{equation}

We estimate that the buffering capacity of this reaction is $\sim3$ mol $\rm H^{+}$/kg/pH unit. Note that $\rm Mg^{2+}$ is expected to behave conservatively in this reaction given the high carbonate alkalinity and high pH of the system, and hence would not produce a large build-up of Mg$^{2+}$ in the Enceladus ocean, in agreement with existing upper limits on Mg \citep{Postberg09}.  Assuming this (or similar) buffering reactions are operative, we can re-calculate the cumulative pH change over 1 billion years. The total acid added over this period (at $2\times10^{10}$ mol/year) is $2\times10^{19}$ moles $H^{+}$. With $3 \times 10^{19}$ kg of water in the ocean, the buffering capacity ($\beta_{Total}$) of the reaction above is:

\begin{equation}
    \beta_{Total} \approx 3 \;\textrm{mol/kg/pH unit}\; \times 3\times10^{19}\;\textrm{kg} \approx 9\times10^{19}\;\textrm{mol/pH unit},
\end{equation}

producing an estimated pH change per 1 billion years of:

\begin{equation}
    \Delta pH \approx \frac{2\times 10^{19}\;\textrm{mol}\;H^{+}}{9\times10^{19}\;\textrm{mol/pH unit}} \approx 0.2\;\textrm{pH units}.
\end{equation}

Together these calculations entail that Enceladus could sustain massive cumulative H$^+$ production without measurable changes to its ocean pH and ocean composition. Existing measurements of either hence provide limited constraints on the possible cumulative O$_2$ and H$_2$ production.

\section{Discussion \label{results-disc}}

\subsection{Comparison of Different \texorpdfstring{H$_2$}{H2} Production Channels on Enceladus}

There are at least three relevant factors to compare when considering different H$_2$-generating processes: possible current production rates, total production potential of H$_2$ during the Enceladus life-time, and the production trajectory over time. A fourth consideration of how different mechanisms may impact the Enceladus water ocean chemistry beyond H$_2$ generation is considered in \S\ref{sec:prebiot}. All these factors are considered by \cite{Bouquet17} for radiolysis and serpentinization, and we hence rely on their analysis for this comparison. 

\begin{figure}[ht]
    \centering
    \includegraphics[width= 0.8\textwidth]{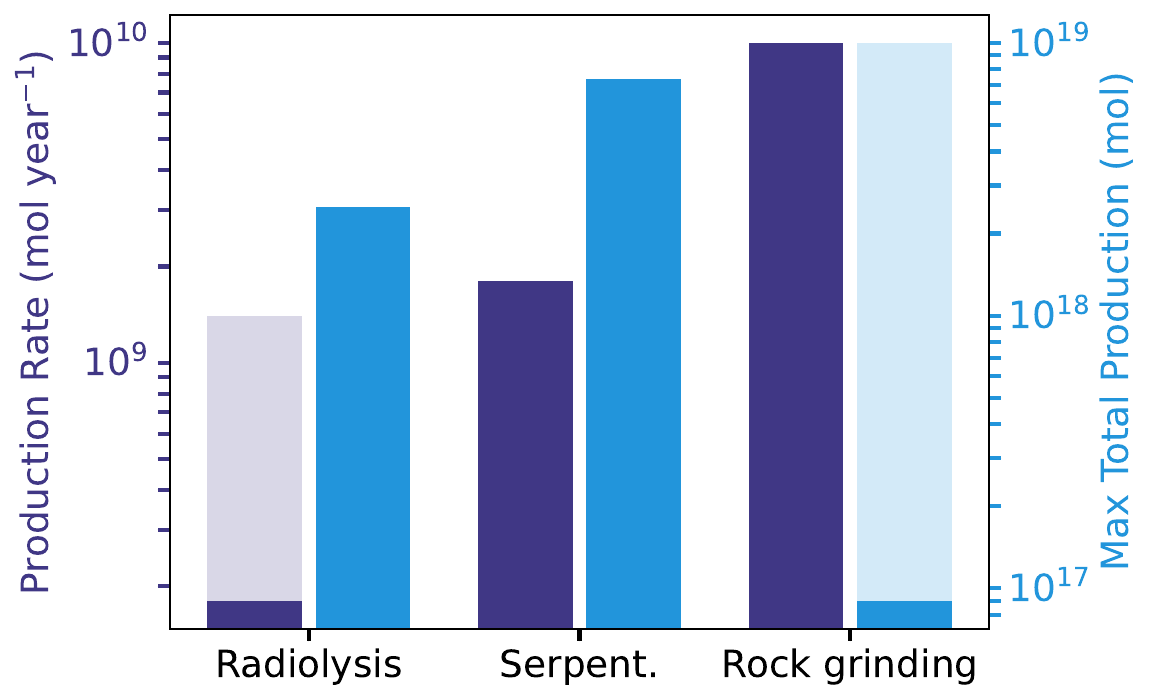}
    \caption{H$_2$ production rates (purple) and total maximum yields (blue) over the lifetime of Enceladus for three different H$_2$  production mechanisms. The radiolysis and serpentinization values come from \cite{Bouquet17}, while the rock grinding results are from this study. The lighter purple for radiolysis indicate the maximum rate, achieved in the young solar system, while the darker purple is at the present time. The lightest blue bar for the rock grinding production channel represents the yield if efficient rock healing mechanisms enable us to release 10\% of the full chemical potential of the Encelaldus core,  while the more intense light blue is based on the maximum yield in rock grinding experiments with no healing \cite{Stone2022}. None of the shown cumulative numbers take into account constraints from Enceladus ocean composition measurements and pH, but as discussed in the text, the buffering capacity of Enceladus entails that existing constraints do not provide strict limitations on the possible H$_2$ production chemistry.}
    \label{fig:comp}
\end{figure}

Figure \ref{fig:comp} shows how the estimates of instantaneous, current production rates compare for radiolysis, serpentization and rock grinding, where the rock grinding value comes from considering energetics only and our fiducial assumptions on how tidal energy is converted into H$_2$ production (\S\ref{sec:model}). This is not a complete apples-to-apples comparison with the radiolysis and serpentinization rates that can be extracted from \cite{Bouquet17} ($\sim1.8\times10^8$ and $\sim1.8\times10^9$ moles year$^{-1}$, respectively, assuming a variable core porosity), which do take into account the progression of both processes over time. However, in the model from \cite{Bouquet17}, H$_2$ production from  serpentinization only varies within a factor of two over the considered time span, and the illustrated rate is hence close to the achieved maximum. In contrast, the H$_2$ production from radiolysis has decreased by an order of magnitude during the lifetime of Enceladus, and  we therefore add a fainter purple bar indicating the H$_2$ production rate from radiolysis in the young Enceladus core, assuming Enceladus is the age of the Solar System. We note that both of these estimates depend on the details of the serpentinization front, which is explored in more detail elsewhere, and here we simply present them as order-of-magnitude comparands to H$_2$ production from tidally induced rock-grinding.

%TC:ignore

%TC:endignore
Figure \ref{fig:comp} shows that the fiducial H$_2$ production value from rock grinding exceeds the most optimistic cases for radiolysis and serpentinization in \cite{Bouquet17} by up to an order of magnitude. However, this fiducial value may be quite optimistic in its own right, since it presumes an efficient conversion of reactive oxygen sites on silicate surfaces into H$_2$ (20\%), and that at least 5\% of the tidal energy deposited in Enceladus results in the production of such oxygen sites. Especially the latter is highly uncertain and would depend on the details of the Enceladus core composition, including the current distribution of rock and pebble sizes, as well as the specifics of the tidal grinding process. More quantitative modeling is needed to confirm whether this order-of-magnitude estimate is indeed plausible. We note, however, that even if only 0.5\% of the tidal energy results in rock grinding and hence reactive O production, rock grinding would be competitive with serpentinization and far exceed the H$_2$ production from radiolysis.

Figure \ref{fig:comp} also compares the expected cumulative H$_2$ production for the three production channels. The estimated cumulative yields from radiolysis and serpentinization from \cite{Bouquet17} are $\sim2.5\times10^{18}$ and $\sim7.4\times10^{18}$ moles, respectively, within a factor of a few of one another. By contrast, the cumulative yield H$_2$ production from rock grinding is more than an order of magnitude lower using our experimentally informed calculation above. This suggests that while rock grinding and the subsequent radical-based H$_2$ production may be responsible for present-day H$_2$ plume measurements, its contribution to the total chemical potential of the Enceladus lifetime would be more modest. This conclusion, however, precludes the possibility of rock healing within Enceladus's core following a rock grinding event. If healing is present the maximum possible yield would instead be set by the chemical potential of the core (\S\ref{sec:model-cum}). To indicate this in Fig. \ref{fig:comp}, we add a faint blue bar representing the total H$_2$ yield corresponding to breaking 10\% of the silicate bonds in the core. 

In addition to presenting different instantaneous production rates and cumulative H$_2$ yields, the three production mechanisms are also expected to evolve differently over time.  This is most straightforward to assess for radiolysis, whose H$_2$ production potential decreases by an order of magnitude over the lifetime of the Solar System due to the decrease in radionuclide abundances. The decrease would be larger still if serpentinization and other erosion processes did not progressively expose a larger portion of the core to water, and hence even this trajectory is highly model dependent \cite{Bouquet17}. The serpentinization rate also decreases with time since the extent of the serpentinization front decreases as it approaches the center of Enceladus, but at a much slower rate \cite{Bouquet17}. Once serpentinization is complete this pathway shuts down, which has been estimated to take a few 100 Myrs \cite{Bouquet17}. Estimating the time evolution of H$_2$ production from rock grinding is more uncertain still, since it depends on the details of the Enceladus core morphology, including the initial and final grain, pebble, and boulder size distribution, the rock grinding efficiency over time, the tidal dynamics etc. Qualitatively it should first increase as more surface area becomes available during the initial core grinding process. After this initial increase it will either begin to decrease as more and more core mass is in the smallest fragments that participate less in the grinding process, or if rock healing is present and efficient, some steady state H$_2$ production may be achieved given a consistent input of tidal energy. Tidal energy input may be far from constant, however \cite[e.g.][]{Goldreich2025}, and an alternative scenario would instead consist of a background of radiolytically generated H$_2$ with spikes of H$_2$ from rock grinding and serpentinization.

To complicate matters further, serpentinization and rock-grinding H$_2$ production channels may interact in rather complex ways. Rock grinding may speed up chemical modification, and chemical modification in aqueous solution (such as serpentinization) may reduce the availability of anhydrous silicate surfaces on the serpentinization timescales. This matters for the current H$_2$ production estimates, since it is currently unclear whether hydrated minerals can produce radical oxygen surfaces when exposed to tidal stresses. Quantifying how repeated abrasion of partially serpentinized material depletes reactive surfaces would be an important target for future work and may be addressed by cyclic grinding–hydration experiments. In the meantime, we are left with considering the different timescales at which these different processes operate. In particular, H$_2$ formation from the dissipation of tidal energy and continuous grinding to expose fresh mineral surfaces is quite fast compared to other timescales.  We may therefore expect an interplay between chemical and physical modification to result in a steady state of fresh surface area production, which does not deviate too much from a scenario where water reactions with fresh rock surfaces are considered in isolation.

Considering the considerable uncertainties associated with H$_2$ production through these different mechanisms, we may ultimately only be able to distinguish between  different H$_2$ scenarios based on what other chemical constraints can be discerned from the composition of the Enceladus plume. We have already considered whether the lack of O$_2$ in the plume could be used to rule out H$_2$ production mechanisms that include O$_2$-co-production and concluded it could not. \cite{Affholder2021} considered the presence of abundant methane as possible evidence for methanogenesis, but evaluating the strength of this evidence is complicated by a lack of experimental data on the interactions between a possible organic-rich ocean and different abiotic H$_2$-generating mechanisms. In the next section we consider some of the chemistry that may be accompanying abiotic H$_2$ production on Enceladus, but acknowledge that experiments are critically needed to make real progress.

\subsection{\texorpdfstring{H$_2$}{H2} Generation and Prebiotic Chemistry on Enceladus \label{sec:prebiot}}

Taking into account the full suite of H$_2$ production channels and their different time scales, a plausible scenario would include a  slowly decreasing background production of H$_2$ over geological timescales through radiolysis, an order of magnitude higher H$_2$ production through serpentinization for a more limited time of hundreds of millions of years, and an even higher H$_2$ production rate from tidally induced rock grinding, but which may only last for millions of years if there is no efficient rock healing. Because of these different timescales, the different processes may play qualitatively different roles on Enceladus, both  when considering its habitability, and the potential to form essential precursors for life.
%In a prebiotic context, H$_2$ may provide the reductive power needed to reduce CO$_2$ to simple organic molecules. 
The relative significance of the different H$_2$ production processes will critically depend on whether organic products once formed can survive over geological times (perhaps stored in ice), or whether they require constant regeneration. If storage of organic molecules is effective, the slow and steady H$_2$ production from radiolysis may have most prebiotic import. Recent models of Titan suggests that even inherited organics can indeed survive up until the present day \cite{Miller2019}, though it is unclear how much of this model translates to Enceladus. By contrast, in the absence of such storage options, the higher instantaneous H$_2$ fluxes achieved with rock grinding may be critical to build up sufficient concentrations of prebiotically interesting molecules in the Enceladus ocean, and to explain the large array of organics observed to originate from the Enceladus plume \cite{Postberg2018}. Which scenario is more realistic is currently speculative, and would highly benefit from experiments on the stability of different organic molecules under Enceladus-like conditions. We also note that there is some evidence for organic matter in the Europa ocean \cite{Trumbo2023}, and hence experiments that explore how the destructive and productive organic chemistries of the two moons cohere or diverge would be especially interesting.

Setting aside the question of timescales, we can use experiments carried out in the context of terrestrial hydrothermal vents to speculate about the role of H$_2$ in the Enceladus prebiotic ecosystem. Such experiments have demonstrated that in the presence of common hydrothermal minerals and elevated temperatures ($\geq$100$^o$C) H$_2$ can reduce inorganic carbon (CO$_2$) to form organic molecules, perhaps providing a prebiotic pathway to the  ancient biological route of CO$_2$ fixation through the acetyl-CoA pathway \cite[e.g.][]{McDermott2015,McCollom2016,Preiner2020}. The effectiveness of this process, especially at lower temperatures, is debated, however, but interestingly \cite{McCollom2016} finds that to reduce CO$_2$ all the way to CH$_4$ requires additional H$_2$ from what can be produced through serpentinization. In the case of Enceladus this could be achieved through rock grinding, and %More generally, if rock grinding can produce large bursts of H$_2$ production, this may enable prebiotic fixation of carbon at levels that could not be achieved through either serpentinization or photolysis. 
experiments directly investigating the organic chemistry that could proceed from H$_2$ generated through rock grinding at a range of temperatures would help to settle this unknown. 

Each of the  H$_2$-generating processes also have the potential to simultaneously produce ROS by splitting water during radiolysis~\cite{pastina1999hydrogen} and  rock grinding~\cite{he2021abiotic}, and perhaps at a lower level through Fenton chemistry during serpentinization. These ROS could play a destructive role, damaging organic precursors formed through H$_2$-powered reduction chemistry, which requires further investigation.  ROS may also, however, have a positive impact on the prebiotic chemistry. H$_2$O$_2$ may aid in bringing key reactants into solution; \cite{Li2016Sulfur} finds evidence that sulfate on the young Earth originated from oxidation of sulfide minerals in the Archaean host rocks by dissolved oxidants such as H$_2$O$_2$. Another oxidation reaction of interest is that of NH$_3$;  \cite{silver2012origin} demonstrated the radiolytic oxidation of NH$_3$ to NO$_3^{-}$ in anaerobic groundwater, and on Enceladus dissolved $\cdot$OH and H$_2$O$_2$ could play a similar role.  Additionally, H$_2$O$_2$ may also act to transform reduced forms of phosphorous to precursors to ATP. H$_2$O$_2$ may hence participate in the production of key ingredients of metabolic cycles \cite{Omran2022}, as well as phosphate, a common limiting nutrient of ecosystems on Earth~\cite[for a review, see][]{duhamel2021phosphorus}. This suggests that radiolytically and geologically-generated oxidant (ROS) and reductant (H$_2$) pools can drive complete C, S, and N cycles. To simultaneously power such reducing and oxidizing reactions of prebiotic or metabolic interest may, however, require the spatial separation of H$_2$ and H$_2$O$_2$.  Partial separation is perhaps achieved by the different diffusion rates of the two molecules as well as their different solubilities in water, their different reactivities, as well as different expected entrapment rates in water ice. These proposed separation mechanisms are speculative in the case of Enceladus, but investigations on H$_2$ production in terrestrial environments have revealed that co-produced oxidants often behave very differently from the H$_2$. In particular \cite{Lin2005} found no excess H$_2$O$_2$ in terrestrial fracture environments with excess H$_2$, even though the proposed radiolytic origin of H$_2$ must have resulted in H$_2$O$_2$ and O$_2$ production. They suggest that the co-produced oxidants are likely quickly consumed through reactions with sulfides, which may also occur on Enceladus.

Finally, when rock grinding results in the production of reactive oxygen on the rock surface, this oxygen could potentially react with other compounds than water. In particular, the Enceladus ocean may contain large amounts of primordial organics compared to terrestrial oceans \cite{Reynard2023}. Carbonaceous chondrites and comets, two possible compositional precursors of Enceladus, are both highly enriched in organic matter, including organic soluble molecules \cite{Mumma11,Alexander03,Altwegg19,Glavin2025}. Silicate oxygen radicals may directly fuel a rich organic chemistry at the ocean-rock interface, where the starting material is not inorganic carbon, but rather pre-existing organic molecules originating in the interstellar medium \cite{Oberg21_Review}.

\subsection{Model Uncertainties and Future Developments}

Many aspects of this "idea" paper are by necessity speculative and here we summarize the main model uncertainties, and outline future theoretical and experimental developments that would put our preliminary conclusions on firmer ground, or challenge them, dependent on the outcome.

The most fundamental unknown for the purpose of this study is the production rate of fresh rock and grain surfaces under the conditions pertinent to the Enceladus core. Detailed models on tidal dissipation mechanisms for different core compositions would aid to retire some uncertainties, but there may also be a need for laboratory experiments on ROS formation when collections of pebbles and grains are exposed to a dynamical environment analogous to a tidally deformed moon core. 

Second, there is a dearth of experiments on H$_2$, O$_2$ and H$_2$O$_2$ generation under conditions that closely mimic expected ocean temperatures and pressures, as well as rock and ocean compositions of Enceladus. We expect that such experiments would result in cumulative yields similar to those of \cite{Stone2022}, but note that this is far from certain, especially when considering that \cite{Stone2022} only carried out a single experiment at the low temperatures that may be most relevant to the the majority of Enceladus rock-ocean interfaces.

Third, whether rock healing is efficient on Enceladus or not radically changes the potential life time production of H$_2$ through rock grinding. We don't know of any models considering rock healing under conditions pertinent to Enceladus and both new theoretical and experimental investigations may be required.

Fourth, we can at present only speculate about possible interactions between the three proposed H$_2$ formation channels. \cite{Bouquet17} showed that the effectiveness of radiolysis increases when taking into account serpentinization because as the serpentinization front moves through the core, it exposes more (radioactive) core material to water. Tidally induced grinding may have a similar effect, and may also speed up serpentinization due to the resulting increasing grain surface area exposed to water with time. Serpentinization may also impact the effectiveness of the rock grinding H$_2$ production channel, since it changes the core composition. Models taking into account all three mechanisms and their interactions are needed as a next step.

Fifth, little is known about the chemical progression of organic-rich oceans in the presence of an efficient H$_2$ (and O$_2$ and H$_2$O$_2$) source, and experiments are needed to explore the possible complex organic chemistry under these conditions. Such experiments could then help define future Enceladus missions, and help distinguish between different sources of H$_2$. 

There are many other model uncertainties as well, including transport efficiencies under different core compositional assumptions, the actual Enceladus core porosity and pebble and grain size distribution, the initial composition of the core and of the ocean etc. Some of these will only be conclusively constrained by direct measurements, while others could at least be better understood through further modeling and constraints on Enceladus's formation following on e.g. \cite{Castillo2018,Neumann2019}.

\section{Summary and Conclusions}

The presence of H$_2$ in the Enceladus plume is intriguing and multiple mechanisms have been proposed to explain it. In this paper we explored H$_2$ production through the interactions between tidally generated fresh rock surfaces, characterized by high concentrations of oxygen radicals, and water. Based on existing experiments, we found that dissipating as little as half a percent of the total estimated tidal energy through rock grinding is sufficient to produce the observed amount of H$_2$.  We thus suggest that rock grinding can contribute a significant instantaneous H$_2$ flux to the Enceladus ocean. Whether this process can be sustained over geological timescales is more uncertain, and critically depends on whether the Enceladus core experiences efficient rock healing.

Compared to the other two most commonly proposed H$_2$ production channels on Enceladus, radiolysis and serpentinization, rock grinding seems to have more potential to produce large instantaneous H$_2$ fluxes, but require more fine-tuned conditions to be active at a high efficiency longer than tens of millions of years. The Enceladus ocean chemistry may therefore be characterized by an interaction of multiple chemical energy sources that operate on different timescales, resulting in the build-up over time of a diverse set of chemical reservoirs. In particular we note that while radiolysis and rock grinding are expected to produce both key reducing and oxidizing agents, serpentinization likely produces only H$_2$. Much theoretical and experimental work remains, however, to both better quantify H$_2$ production on Enceladus, and to work out its prebiotic consequences.  

\section*{Acknowledgement}

This article was conceived of during a series of meetings of the Origins Federation, and we are grateful to the four Origins Federation institutes (the Harvard Origins of Life Initiative, the ETH Centre for Origin and Prevalence of Life, the Leverhulme Center for Life in the Universe at University Cambridge, and the University of Chicago Center for the Origins of Life), as well as an anonymous donor who hosted one of the meetings. This work was supported by a grant from the Simons Foundation 686302, KIÖ.

\bibliographystyle{elsarticle-num}
%\bibliography{mybib_epsl.bib}





% \end{thebibliography}

\end{document}